\documentclass[preprint,aps,showpacs,prb,floatfix]{revtex4}

\usepackage{graphicx}
\usepackage{dcolumn}
\usepackage{bm}

\begin{document}


\title{
Interaction energy functional for lattice density functional theory:\\
Applications to one-, two- and three-dimensional Hubbard models\\ 
}

\author{R.~L\'opez-Sandoval} 
\affiliation{Instituto Potosino de Investigaci\'on Cient\'{\i}fica
y Tecnol\'ogica,
Camino a la presa San Jos\'e 2055, 78216 San Luis Potos\'{\i}, M\'exico}
\author{G.~M.~Pastor}
\affiliation{Laboratoire de Physique Quantique, 
Centre National de la Recherche Scientifique,
Universit\'e Paul Sabatier, 31062 Toulouse, France}

\date{\today}

\begin{abstract}  

The Hubbard model is investigated in the framework of lattice 
density functional theory (LDFT). The single-particle density 
matrix $\gamma_{ij}$ with respect the lattice sites is considered
as the basic variable of the many-body problem. A new
approximation to the interaction-energy functional 
$W[\gamma]$ is proposed which is based on its scaling properties
and which recovers exactly the limit of strong electron correlations 
at half-band filling.
In this way, a more accurate description of $W$ is obtained 
throughout the domain of representability of $\gamma_{ij}$, 
including the crossover from weak to strong correlations.
As examples of applications results are given for the 
ground-state energy, charge-excitation gap, 
and charge susceptibility of the Hubbard model in one-, two-, 
and three-dimensional lattices. The performance of the method
is demonstrated by comparison with available exact solutions,
with numerical calculations, and with LDFT using a simpler 
dimer ansatz for $W$. Goals and limitations of the different 
approximations are discussed. 

\end{abstract}


%
\maketitle

\section{\label{sec:introd}
Introduction}

Density functional theory (DFT) provides a rigorous framework for 
studying the physics of the many-body 
problem.\cite{hk,ks,parr-book,gross-book}
The fundamental concept behind DFT is to replace the conventional
dynamical variables that completely define a many-particle state 
(e.g., the wave function in a quantum mechanical problem, or 
the particle positions and momenta in a classical system) 
by considering the particle-density distribution $\rho(\vec r)$ 
as the basic variable. 
For this purpose the energy $E$ of the system is expressed as a 
functional of $\rho(\vec r)$ separating the various energy terms
in two main groups. The first one contains the contributions that
depend explicitly on the problem under study, which result from the
coupling between $\rho(\vec r)$ and the external fields $V_{ext}(\vec r)$. 
The second one concerns the intrinsic energy of the many-particle system, 
namely, the kinetic energy $T$ and the interaction energy $W$
of the particles. These terms are universal functionals of 
$\rho(\vec r)$ in the sense that they are independent of the problem 
under study. While the general functional dependence 
of $T[\rho]$ and $W[\rho]$ is not known explicitly, they 
can be formally expressed as the result 
of integrating out the microscopic degrees of freedom.
In the case of ground-state electronic properties, this is achieved 
by imposing that $T[\rho]$ and $W[\rho]$ correspond to the minimum possible 
value of $T+W$ for a given $\rho(\vec r)$.\cite{parr-book,gross-book,levy}
These basic notions have general validity 
and are therefore relevant to a wide variety of situations 
which may have very different physical origins. 

In the present paper the concepts of DFT are applied to investigate
the physics of strongly correlated electrons in a narrow energy band.
The theoretical description of these systems is usually 
based on lattice Hamiltonians such as Anderson,\cite{andmod} 
Hubbard,\cite{hub} Pariser-Parr-Pople,\cite{ppp} and related 
models which focus on the most relevant electron dynamics at low 
energies.\cite{mb-books,dagotto,georges,imada}
The study of many-body lattice models in the framework of DFT seems
particularly interesting from various perspectives. On the one side, 
a detailed understanding of the electronic properties in the 
strongly correlated limit constitutes an important 
theoretical challenge. Exact results are rare or numerically 
very demanding and a variety of elaborate many-body techniques are 
specifically developed for their study. Therefore,
a density functional approach with an appropriate ansatz for $W$
should be a useful alternative tool for investigating at least some
aspects of this complex problem. On the other side, 
one would like to extend the range of applicability of DFT to 
strongly correlated phenomena, like the separation of charge and spin 
degrees of freedom or the correlation induced localization, where conventional 
local density approximations (LDAs) or generalized gradient approximations
(GGAs) are known to fail systematically. Moreover, the development 
of DFT on a lattice constitutes an intrinsically inhomogeneous approach, 
which provides a true alternative to methods relying on the homogeneous
electron gas. In this context, investigations on
many-body models should open new insights into the
properties of the interaction-energy functional that could also 
be useful for future extensions to more realistic Hamiltonians 
and first principles calculations.

In past years, a number of density functional studies
of lattice models have performed concerning in particular
the determination of band-gaps in semiconductors,\cite{gunn} 
the role of off-diagonal elements of the density matrix 
and the non-interacting $v$ representability in strongly 
correlated systems,\cite{godby} or the development of energy 
functionals of the density matrix with applications to Hubbard 
and Anderson models.\cite{carl} In previous papers we have formulated
a lattice density functional theory (LDFT) of many-body models 
by considering the density matrix $\gamma_{ij}$ with respect to 
the lattice sites $i$ and $j$ as the fundamental 
variable.\cite{ldftxcfun,ldftscfer,ldftdim1D}
The interaction energy $W$ of the Hubbard  
model has been calculated exactly as a function of 
$\gamma_{ij}$ for various periodic lattices having 
$\gamma_{ij} =\gamma_{12}$ for nearest neighbors (NNs) $i$ and $j$. 
On this basis, a simple general approximation 
to $W(\gamma_{12})$ has been proposed which derives from exact dimer 
results, the scaling properties of $W$, and known limits. 
Using this ansatz, several ground-state properties
of one-dimensional (1D) and two-dimensional (2D) systems have 
been obtained in good agreement with available exact solutions
and accurate numerical calculations.\cite{ldftscfer} In addition,
applications to dimerized chains provided us with systematic 
results for the ground-state energy and charge excitation gap
of the 1D Hubbard model as a function of hopping alternation
and Coulomb repulsion, including the crossover from weak to strong 
correlations.\cite{ldftdim1D} LDFT appears therefore as an efficient 
method of determining the electronic properties of many-body lattice 
models, thus encouraging further developments and applications.

The main purpose of this paper is to present a new approximation
to the interaction-energy functional $W[\gamma]$ of the Hubbard 
Hamiltonian and to apply it to determine several electronic properties 
of this model in the framework of LDFT. In Sec.~\ref{sec:ldft} 
the basic formulation of LDFT is briefly reviewed. 
Different approximations to interaction-energy functional 
are presented and discussed
in Sec.~\ref{sec:W}. First, we analyze the properties of a 
previously proposed dimer ansatz\cite{ldftscfer} and discuss its 
goals and limitations by comparison with known exact results. 
Some shortcomings of this functional in the limit of strong 
correlations at half-band filling are pointed out. In order to 
overcome them, we propose a new approximation  
based on the scaling properties of $W$, which recovers the
correct behavior in the limit of strong interactions. In this way 
a more accurate description of the dependence on $\gamma_{ij}$ 
in different dimensions and lattice structures is obtained. 
The following sections are mainly concerned 
with applications to 1D, 2D and 3D Hubbard models. 
Results for the ground-state energy, charge-excitation gap, 
and charge susceptibility are presented in 
Secs.~\ref{sec:Egs}, \ref{sec:gaps}, and \ref{sec:chsus}, respectively.
Comparison is made with the simpler dimer ansatz and with exact 
analytical or numerical solutions, whenever available, in order 
quantify the accuracy of the different approximations.
Finally, Sec.~\ref{sec:concl} summarizes the conclusions
and points out some perspectives of future developments.

\section{\label{sec:ldft}
Density-functional theory of lattice models}

In order to be explicit we focus on the Hubbard model which is expected 
to capture the main physics of lattice fermions in a narrow energy 
band. The Hamiltonian\cite{hub}
\begin{equation}
\label{eq:hamhub}
H = \sum_{\langle i,j\rangle \sigma} t_{ij}
\hat c^{\dagger}_{i \sigma} \hat c_{j \sigma} +
U \sum_i  \hat n_{i \downarrow} \hat n_{i\uparrow} ,
\end{equation}
includes nearest neighbor (NN) hoppings $t_{ij}$, and on-site 
interactions given by $U$  ($\hat n_{i\sigma} = 
\hat c_{i\sigma}^\dagger \hat c_{i\sigma}$).  
The hopping integrals $t_{ij}$ are defined by the lattice structure
and by the range of the single-particle hybridizations
(typically, $t_{ij} = - t < 0$ for NN $ij$). They specify
the system under study and thus 
play the role given in conventional DFT to the external 
potential $V_{ext}(\vec r)$. 
Consequently, the basic variable in LDFT is 
the single-particle density matrix $\gamma_{ij}$.
The situation is similar to the density-matrix functional
theory proposed by Gilbert for the study of nonlocal  
pseudopotentials,\cite{gilb} since the hoppings are nonlocal in the sites.
A formulation of DFT on a lattice in terms of the diagonal 
$\gamma_{ii}$ alone is possible only if one restricts oneself to models 
with constant $t_{ij}$ for $i\not= j$. In this case the functional 
$W[\gamma_{ii}]$ depends on the values of $t_{ij}$ for $i\not= j$ 
and in particular on $U/t$.\cite{gunn}

The ground-state energy $E_{gs}$ and density-matrix $\gamma_{ij}^{gs}$
are determined by minimizing the energy functional  
\begin{equation}
\label{eq:E}
E[\gamma] = E_K[\gamma] + W [\gamma]
\end{equation}
with respect to $\gamma_{ij}$. $E[\gamma]$ is defined for all 
density matrices that derive from a physical state, i.e., that can be 
written as
\begin{equation}
\label{eq:gamij} 
\gamma_{ij} = \sum_{\sigma}\gamma_{ij \sigma}=
\sum_\sigma \langle \Psi| \hat c_{i \sigma }^{\dagger}
                          \hat c_{j \sigma } |\Psi \rangle \; ,
\end{equation}
where $|\Psi \rangle$ is an $N$-particle state. 
Such $\gamma_{ij}$ are said to be pure-state $N$-representable.
An extension of the definition domain of $E$ to 
ensemble-representable density matrices $\Gamma_{ij}$ is straightforward 
following the work by Valone.\cite{val,foot_ens} 
The first term in Eq.~(\ref{eq:E}) is the kinetic energy 
associated with the electronic motion in the lattice. It is given by
\begin{equation}
\label{eq:EK}
E_K[\gamma] = \sum_{ij} t_{ij} \gamma_{ij} \; ,
\end{equation}
thus including all single-particle contributions.  
The second term is the interaction-energy functional
given by\cite{levy}
\begin{equation}
\label{eq:Wex}
W[\gamma] = 
{\min_{\Psi\to\gamma}} \left[ U \sum_i  \langle \Psi [\gamma] |
\hat n_{i\uparrow} \hat n_{i\downarrow}|
\Psi [\gamma] \rangle  \right] \; ,
\end{equation}
where the minimization runs over all 
$N$-particles states $| \Psi [\gamma] \rangle$ 
that satisfy 
\begin{equation}
\label{eq:rep}
\langle \Psi [\gamma] | \; 
\sum_\sigma \hat c_{i \sigma }^{\dagger}\hat c_{j \sigma} \; 
|\Psi [\gamma] \rangle = \gamma_{ij} 
\end{equation}
for all $i$ and $j$. $W[\gamma]$ represents the 
minimum value of the interaction energy compatible with a given 
density matrix $\gamma_{ij}$.
It is a universal functional of $\gamma_{ij}$ in 
the sense that it is independent of $t_{ij}$, i.e., of the 
system under study. However, $W$ depends on the number of electrons $N_e$,
on the structure of the many-body Hilbert space, as given by 
$N_e$ and the number of orbitals or sites $N_a$, and on the 
form of the model interactions.\cite{foot_gen} 

$E[\gamma]$ is minimized by expressing
\begin{equation} 
\gamma_{ij} = \sum_\sigma \gamma_{ij\sigma}
=\sum_{k\sigma} u_{ik\sigma} \eta_{k\sigma} u_{jk\sigma}^*
\end{equation} 
in terms of the eigenvalues $\eta_{k\sigma}$ (occupation 
numbers) and eigenvectors $u_{ik\sigma}$ (natural orbitals) of
$\gamma_{ij\sigma}$.
Lagrange multipliers $\mu$ and $\lambda_{k\sigma}$ 
($\varepsilon_{k\sigma} = \lambda_{k\sigma} / \eta_{k\sigma}$) are
introduced in order to 
impose the constraints $\sum_{k\sigma} \eta_{k\sigma} = N_e$ and
$\sum_i\vert u_{ik\sigma}\vert^2 = 1$. Deriving with 
respect to $u_{jk\sigma}^*$ and $\eta_{k\sigma}$
($0 \le \eta_{k\sigma}\le 1$), one obtains the eigenvalue 
equations\cite{ldftscfer,gilb}
\begin{equation}
\label{eq:minsc}
\sum_i \left(t_{ij} + {\partial W \over \partial \gamma_{ij\sigma}} \right)
u_{ik\sigma} = \varepsilon_{k\sigma} u_{jk\sigma} \; ,
\end{equation}
with the subsidiary conditions
$\varepsilon_{k\sigma} < \mu$ if $\eta_{k\sigma} = 1$,
$\varepsilon_{k\sigma} = \mu$ if $0 < \eta_{k\sigma} < 1$,
and 
$\varepsilon_{k\sigma} > \mu$ if $\eta_{k\sigma} = 0$.
Self-consistency is implied by the dependence of
$\partial W / \partial\gamma_{ij\sigma}$ on $\eta_{k\sigma}$ and
$u_{ik\sigma}$. This formulation is analogous to density-matrix 
functional theory in the continuum.\cite{gilb} 
However, it differs
from KS-like approaches which assume non-interacting 
$v$-representability and where only integer occupations 
are allowed.\cite{gunn,godby}
In the present case, the fractional occupations 
of natural-orbitals play a central role. One may in fact show
that in general $0<\eta_{k\sigma}<1$ 
for all $k\sigma$. Exceptions are found in very special situations like the 
uncorrelated limit ($U = 0$) or the fully-polarized ferromagnetic 
state in the Hubbard model ($S_z = \min\{N_e, 2N_a-N_e\} / 2$). 
This can be understood from perturbation-theory 
arguments ---none of the $\eta_{k\sigma}$ is a
good quantum number for $U\not= 0$--- and is explicitly verified 
by exact solutions of the Hubbard Hamiltonian on finite systems or the
1D chain.\cite{lieb-wu} Therefore, all $\varepsilon_{k\sigma}$
in Eq.~(\ref{eq:minsc}) must be degenerate and consequently the
ground-state density matrix satisfies
\begin{equation}
\label{eq:tij}
t_{ij} + {\partial W \over \partial\gamma_{ij\sigma}} 
= \delta_{ij} \; \mu \;.
\end{equation}
Notice the importance of the dependence of $W$ on the off-diagonal 
density-matrix elements $\gamma_{ij}$ which measure the 
degree of electron localization.
Approximations of $W$ in terms of the 
diagonal $\gamma_{ii}$ alone are not applicable in this 
framework ($t_{ij}\not= 0$ for NN $ij$). Eq.~(\ref{eq:tij})
provides a self-consistent scheme to obtain the ground-state
$\gamma_{ij}^{gs}$ according to the variational principle. 
In the following section we present and discuss simple explicit
approximations to 
$W[\gamma]$ that are intended to describe the electronic 
properties of the Hubbard model in different interaction
regimes, band-fillings, and lattice structures.

\section{\label{sec:W}
Interaction-energy functional in the Hubbard model}

The general functional $W[\gamma]$, valid for all lattice structures 
and for all types of hybridizations, can be simplified at the expense of 
universality if the hopping integrals are short ranged. For example, 
if only NN hoppings are considered, the kinetic energy $E_K$ is 
independent of the density-matrix elements between sites that are not NNs. 
Therefore, the constrained search in Eq.~(\ref{eq:Wex}) may be restricted 
to the $| \Psi [\gamma] \rangle$ that satisfy Eq.~(\ref{eq:rep})
only for $i=j$ and for NN $ij$. This reduces significantly the number 
of variables in $W[\gamma]$ and renders the determination and
interpretation of the functional dependence far simpler. 
In particular for periodic lattices the ground-state 
$\gamma_{ij}^{gs}$ is translational invariant. Therefore, 
in order to determine $E_{gs}$ and 
$\gamma_{ij}^{gs}$, one may set $\gamma_{ii} = n = N_e/N_a$ for all 
sites $i$, and $\gamma_{ij} = \gamma_{12}$ for all NN pairs $ij$. 
In this case the interaction energy can be regarded as a simple function 
$W(\gamma_{12})$ of the density-matrix element between NNs. 
This is certainly a great practical advantage. However, it should 
be noted that restricting the minimization constraints 
in Eqs.~(\ref{eq:Wex}) and (\ref{eq:rep}) 
to NN $\gamma_{ij}$ also implies that $W$ loses its 
universal character, since the NN map and the resulting dependence 
of $W$ on $\gamma_{12}$ are in principle different for different 
lattice structures.

The difficulties introduced by the lack of universality 
can be overcome by taking advantage of the scaling 
properties $W(\gamma_{12})$. Recent numerical studies\cite{ldftxcfun}  
have in fact shown that $W$ is nearly independent of system size
$N_a$, band filling $n=N_e/N_a$ and lattice structure, if 
W is measured in units of the Hartree-Fock energy 
$E_{\rm HF} = U n^2 / 4$ 
and if $\gamma_{12}$ is scaled within the relevant 
domain of representability $[\gamma_{12}^\infty, \gamma_{12}^0]$.
Here, $\gamma_{12}^0$ stands for the largest possible 
value of the NN bond order 
$\gamma_{12}$ for a given $N_a$, $n$, and lattice structure. It 
represents the maximum degree of electron delocalization and
corresponds to the uncorrelated limit. 
On the other side, $\gamma_{12}^\infty$ refers to the strongly 
correlated limit of $\gamma_{12}$, i.e., to the largest NN bond 
order that can be obtained under the constraint of vanishing $W$.
For half-band filling $\gamma_{12}^\infty =0$, 
while for $n \not= 1$, $\gamma_{12}^\infty>0$.\cite{nonbip}
Physically, the possibility of scaling the interaction energy
means that the relative change in $W$ associated to a 
given change in the degree of electron localization 
$g_{12} = (\gamma_{12}    - \gamma_{12}^{\infty}) / 
          (\gamma_{12}^{0}-\gamma_{12}^{\infty})$ can be regarded 
as nearly independent of the system under study. 
This pseudo-universal behavior of $W/E_{\rm HF}$ as
a function of $g_{12}$ can be exploited to 
obtain good general approximations to $W(\gamma_{12})$
by applying such a scaling to the functional dependence 
derived from a simple reference system or from known limits. 

In a previous paper we have proposed an approximation to 
the interaction energy $W$ of the Hubbard model by 
extracting the functional dependence from the exact result 
for the Hubbard dimer, which is given by\cite{ldftscfer}
\begin{equation}
W^{(2)} = E_{\rm HF} 
\left( 1 - \sqrt{1 - g_{12}^2 } \right) \; .
\label{eq:W2}
\end{equation}
This very simple expression satisfies several general properties
of the exact $W(\gamma_{12})$:\\
(i) For $\gamma_{12} = \gamma_{12}^0$, $W^{(2)} = E_{\rm HF}$ since 
the underlying electronic state $\Psi[\gamma_{12}^0]$ is a 
single Slater determinant. Moreover, one observes that  
$\partial W^{(2)} / \partial \gamma_{12}=\infty$ for 
$\gamma_{12} = \gamma_{12}^0$. This is a necessary
condition in order that $\gamma_{12}^{gs} < \gamma_{12}^0$
already for arbitrary small $U/t \not= 0$, 
as expected from perturbation theory. \\
(ii) $W^{(2)}(\gamma_{12})$ decreases monotonously
with decreasing $\gamma_{12}$ reaching 
its lowest possible value, $W=0$, for $\gamma_{12} = \gamma_{12}^\infty$. 
In other words, a reduction of the interaction energy is obtained
at the expense of electron delocalization. \\
(iii) In the strongly correlated limit ($\gamma_{12}\ll  \gamma_{12}^0$)
one observes that $W^{(2)} \propto \gamma_{12}^2$. Therefore, 
for $U/t\gg 1$, $\gamma^{gs} \propto t/U$ and $E_{gs} \propto t^2/U$, 
a well known result in the Heisenberg limit of the half-filled Hubbard 
model.\cite{mb-books}\\
A correct description of these basic properties and of the 
dependence of $W/E_{HF}$ on $g_{12}$ are at the origin of 
the remarkable performance of this simple dimer ansatz
in the description of several ground-state properties 
of the Hubbard model.\cite{ldftscfer}

In order to discuss the strongly correlated limit of Eq.~(\ref{eq:W2})
in more detail we expand $W^{(2)}$ to lowest order 
in $\gamma_{12}$. At half-band filling one obtains
\begin{equation}
W^{(2)} = (1/8) \alpha_2 U \gamma_{12}^2 + {\cal O}(\gamma_{12}^4)
\label{eq:W2inf}
\end{equation}
with $\alpha_2 = (\gamma_{12}^0)^{-2}$.
The exact interaction-energy $W_{ex}$ is also proportional 
to $U \gamma_{12}^2$ in the limit of small $\gamma_{12}$.
Therefore, $W_{ex}$ can be expanded in the same form as Eq.~(\ref{eq:W2inf})
but with a somewhat different coefficient $\alpha_{ex}$.
Notice that in the case of the Hubbard dimer, we have $\gamma_{12}^0 = 1$ 
and $\alpha_2^{dim} = 1$, which coincides of course with the exact result.
Considering for example the 1D chain, the 2D square lattice, 
and the 3D simple-cubic lattice one finds that the leading
coefficients resulting from Eq.~(\ref{eq:W2}) are
$\alpha_2^{1D} = (\pi/2)^2 \simeq 2.47$,
$\alpha_2^{2D} = 6.09$, and
$\alpha_2^{3D} = 9.30$. 
These can be compared with the corresponding exact result
derived from the Bethe-ansatz solution of the 1D Hubbard chain,\cite{lieb-wu} 
or with perturbation-theory calculations for 
the square and simple-cubic lattices,\cite{taka}
which are given by 
$\alpha_{ex}^{1D} = 2/\ln 2 \simeq 2.89$,
$\alpha_{ex}^{2D} = 6.91$, and
$\alpha_{ex}^{3D} = 10.94$. 
One observes that Eq.~(\ref{eq:W2}) reproduces correctly the
trends in $\alpha$ with increasing dimensions.
However, there is also a systematic underestimation 
of the interaction energy of the order of $12$--$15$\%.
These quantitative discrepancies have direct 
consequences on the predicted properties, since
the behavior of $W$ for small $\gamma_{12}$ determines the 
ground-state density matrix $\gamma_{12}^{gs}$ and energy $E_{gs}$ 
in the strongly correlated limit. In fact, approximating $W$
as in Eq.~(\ref{eq:W2inf}), writing the kinetic energy
as $E_K = zt\gamma_{12}$, where $z$ is the coordination number,
and using Eq.~(\ref{eq:tij}), one obtains 
$\gamma_{12}^{gs} = (4z/\alpha)(t/U)$ and
$E_{gs} = -(2z^2/\alpha)(t^2/U)$. 
Thus, an inaccuracy in $\alpha$ results in 
a similar relative error in $\gamma_{12}^{gs}$ and $E_{gs}$ for
$U/t \gg 1$. 

To overcome these shortcomings more flexible approximations to
the interaction-energy functional are needed, which allow one to go beyond
Eq.~(\ref{eq:W2}). Therefore, we propose a general ansatz of the form
\begin{equation}
W^{(n)} = E_{\rm HF} 
\left( 1 - \sqrt{P_n(g_{12})} \right) \; ,
\label{eq:Wn}
\end{equation}
where $P_n(g_{12})$ is a function of 
$g_{12}= (\gamma_{12}    - \gamma_{12}^{\infty}) / 
         (\gamma_{12}^{0}-\gamma_{12}^{\infty})$, 
thus incorporating the scaling properties of $W$ without 
loss of generality. $P_n(g_{12})$ is approximated by an 
$n$-order polynomial $P_n(g_{12}) = \sum_{k=0}^n a_k g_{12}^k$. 
This is justified by the fact that $(W - E_{HF})^2$ is in general
a well-behaved function of $g_{12}$, even in the uncorrelated limit
where $\partial W / \partial \gamma_{12}$ diverges 
($g_{12} = 1$). The coefficients $a_k$ are to be determined 
from known properties of $W$. First of all, one observes that 
at half-band filling, and for bipartite lattices in general, 
the sign of $\gamma_{12}$ can be changed without altering $W$.
Thus, $P_n(g_{12})$ is an even function of $g_{12}$
and $a_k = 0$ for odd $k$ ($\gamma_{12}^{0-}= -\gamma_{12}^{0+}$ 
and $\gamma_{12}^{\infty -} = -\gamma_{12}^{\infty +}$).
In non-bipartite lattices away from half-band filling one 
may also set for simplicity $a_k = 0$ for odd $k$, since 
the dependence on $g_{12}$ is very similar for positive 
and negative $\gamma_{12}$, once the different domains 
of representability are scaled.\cite{ldftxcfun,nonbip}

The uncorrelated and fully-correlated limits of $W$
($W = E_{HF}$ for $g_{12} = 1$, and $W = 0$ for $g_{12} = 0$)
impose two simple conditions on the $a_k$, namely, 
$P_n(1) = \sum_k a_k = 0$ and $P_n(0) = a_0 = 1$. 
This defines the second-order approximation $W^{(2)}$
completely. In this case, $P_2(g_{12}) = 1 - g_{12}^2$, which coincides
with the above discussed dimer ansatz [Eq.~(\ref{eq:W2})]. 
The two approaches are therefore consistent.
The dimer ansatz can be regarded as the simplest polynomial-based
approximation of $W$, as given by Eq.~(\ref{eq:Wn}), 
that satisfies the obvious limits. 

The 4th-order approximation $W^{(4)}$ introduces the aimed additional 
flexibility that can be exploited to reproduce the strongly
correlated limit of $W$ exactly. Expanding Eq.~(\ref{eq:Wn}) 
to second-order in $\gamma_{12}$ one observes that at half-band 
filling this is achieved when $a_0 = 1$, 
$a_2 = - \alpha_{ex} (\gamma_{12}^0)^{2} 
     = - \alpha_{ex} / \alpha_2$, and $a_4 = - (a_0 + a_2)$.
Thus, the 4th-order approximation to $W$ is given by
\begin{equation}
W^{(4)} = E_{\rm HF} 
\left( 1 - \sqrt{1 - \kappa g_{12}^2 + (\kappa -1) g_{12}^4} \right) \; ,
\label{eq:W4}
\end{equation}
where $\kappa = \alpha_{ex} / \alpha_2 > 0$ is the ratio between the 
small-$\gamma_{12}$ expansion coefficients of $W^{ex}$ and $W^{(2)}$.
The value of $\kappa$ depends on the lattice structure
or system dimensions. At half-band filling it can be determined 
by applying perturbation-theory to the 
Heisenberg limit of the Hubbard model.\cite{taka} For instance, 
for the 1D chain, 2D square lattice, and 3D simple-cubic lattice
one obtains, respectively, 
$\kappa_{1D} = 8/(\pi^2\ln 2) = 1.169$,  
$\kappa_{2D} = 1.135$, and  
$\kappa_{3D} = 1.176$. Notice that $\kappa$ depends rather weakly
on the lattice structure and that it is not very far from the
dimer value $\kappa = 1$, for which Eq.~(\ref{eq:W4}) reduces 
to Eq.~(\ref{eq:W2}). Therefore, the 4th-order term
appears as a relatively small correction to the 2nd-order 
approximation. Higher-order polynomial approximations to $W$
could be derived in an analogous way, provided that reliable information
is available on the following terms of the small-$\gamma_{ij}$
expansion of $W^{ex}$. This gives the possibility of further improving the 
accuracy of the results by incorporating a more detailed description of 
the strongly correlated limit.

\begin{figure}
\includegraphics[scale = 0.4]{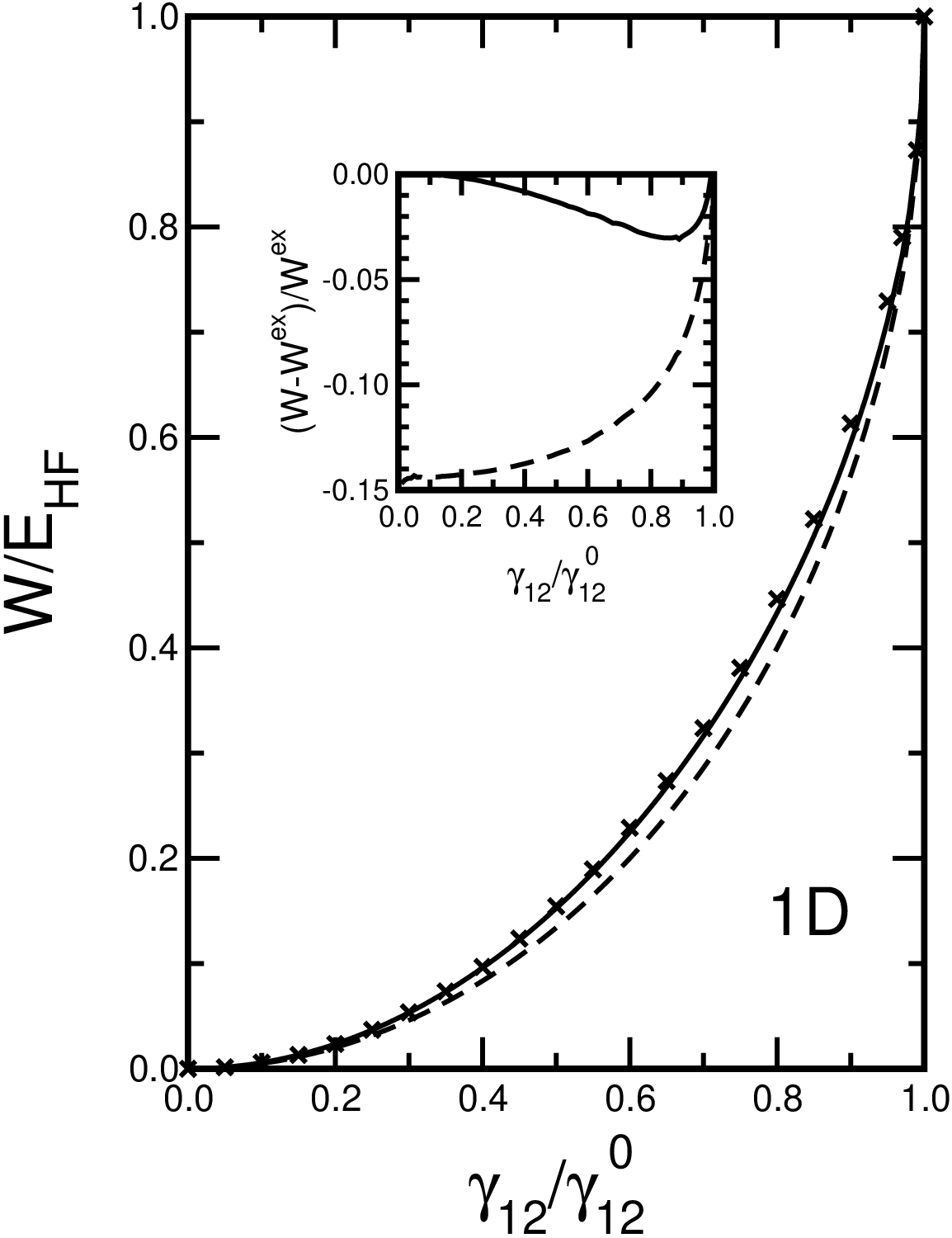}
\caption{\label{fig:W1D}
Interaction energy $W(\gamma_{12})$ of the one-dimensional (1D) 
Hubbard model at half-band filling ($n=1$) as a function of the 
density-matrix element or bond order $\gamma_{12}$ between nearest 
neighbors. $\gamma_{12}^0$ refers to the ground-state bond order 
in the uncorrelated limit ($U=0$). Results are given for
the dimer approximation $W^{(2)}$ [Eq.~(\ref{eq:W2}), dashed],
the 4th-order approximation $W^{(4)}$ 
[Eq.~(\ref{eq:W4}) with $\kappa = \kappa_{1D} = 1.169$, solid],
and the exact $W^{ex}$ [Eq.~(\ref{eq:Wex}), crosses] 
which is derived from the Bethe ansatz solution.\protect\cite{lieb-wu}  
In the inset the corresponding relative errors are shown.
        }
\end{figure}

In Fig.~\ref{fig:W1D} the approximate interaction energies
$W^{(2)}$ and $W^{(4)}$ of the half-filled 1D Hubbard chain 
are compared with the corresponding exact result
$W_{ex}$, as derived from the Bethe-Ansatz solution.\cite{lieb-wu} 
As already observed,\cite{ldftscfer} even the simplest dimer
ansatz $W^{(2)}$ follows $W_{ex}(\gamma_{12})$ quite closely all along 
the crossover from weak to strong correlations.
In this case the interaction energy is always underestimated,
and  the absolute value of the relative error 
$\epsilon = |W - W^{ex}|/W^{ex}$
increases monotonously as $\gamma_{12}$ decreases, reaching about 15\% 
for $\gamma_{12}/\gamma_{12}^0 < 0.4$. The 4th-order approximation 
provides a significant advance, not only for 
$\gamma_{12}/\gamma_{12}^0\ll 1$ but in the complete domain of 
representability. For $W^{(4)}$ the relative error $\epsilon$ 
is reduced to less than 1\% 
for $\gamma_{12}/\gamma_{12}^0 < 0.4$
($\epsilon \to 0$ for $\gamma_{12} \to 0$). 
The largest discrepancies are found for 
$\gamma_{12}/\gamma_{12}^0 \simeq 0.8$--$0.9$, 
where $\epsilon$ reaches only 3\%. An appreciable improvement in 
the accuracy of the derived properties can be therefore expected.
In the following sections, Eqs.~(\ref{eq:W2}) and (\ref{eq:W4}) 
are applied in the framework of LDFT to determine several 
electronic properties of the Hubbard model in 1D, 2D, and 3D
periodic lattices.

\section{\label{sec:Egs}
Ground state energy} 

\begin{figure}
\includegraphics[scale = 0.4]{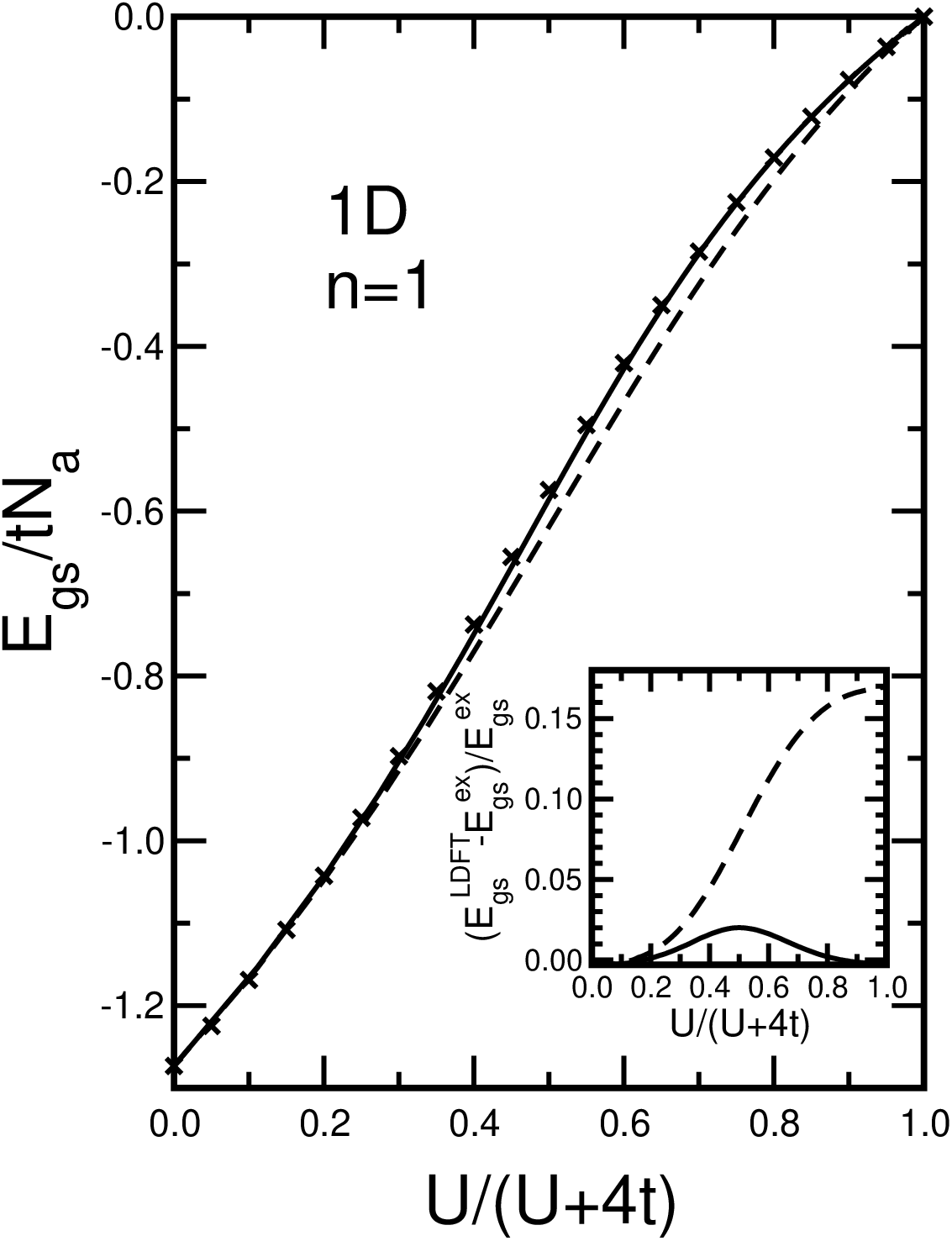}
\caption{\label{fig:Egs1D}
Ground-state energy $E_{gs}$ of the half-filled 1D Hubbard model 
as a function of the Coulomb repulsion strength $U/t$. 
The dashed curves refer to lattice density-functional theory (LDFT)
using the dimer approximation to $W$ [Eq.~(\ref{eq:W2})] and 
the solid curves to the 4th-order approximation 
[Eq.~(\ref{eq:W4}) with $\kappa = \kappa_{1D} = 1.169$].  
The crosses are the exact results derived from the 
Bethe-ansatz solution.\protect\cite{lieb-wu} 
The corresponding relative errors are given in the inset.
        }
\end{figure}

In Fig.~\ref{fig:Egs1D} the ground-state energy $E_{gs}$ of 
the half-filled 1D Hubbard model is given as a function of 
the Coulomb repulsion strength $U/t$. Comparison between
LDFT and the Bethe-Ansatz exact solution shows that 4th-order 
approximation improves significantly the already good
results derived using the dimer ansatz.\cite{ldftscfer}
This concerns not only the strongly correlated limit
where, as expected, $W^{(4)}$ recovers the exact result, but 
the complete range of $U/t$. The largest quantitative 
discrepancies between exact and 4th-order results are 
in fact very small. They amount to less than 3\%  
and are found for intermediate interaction strengths ($U/t\simeq 4$). 
In contrast, the relative error in the dimer ansatz increases 
monotonously with $U/t$ reaching about 17\% for $U/t =\infty$
(see the inset of Fig.~\ref{fig:Egs1D}). It is interesting 
to note that in both cases no artificial symmetry breaking is required 
in order to describe correctly the electron localization
induced by correlations and the resulting dependence of $E_{gs}$ on 
$U/t$, as it is often the case in other mean-field approaches.

\begin{figure}
\includegraphics[scale = 0.4]{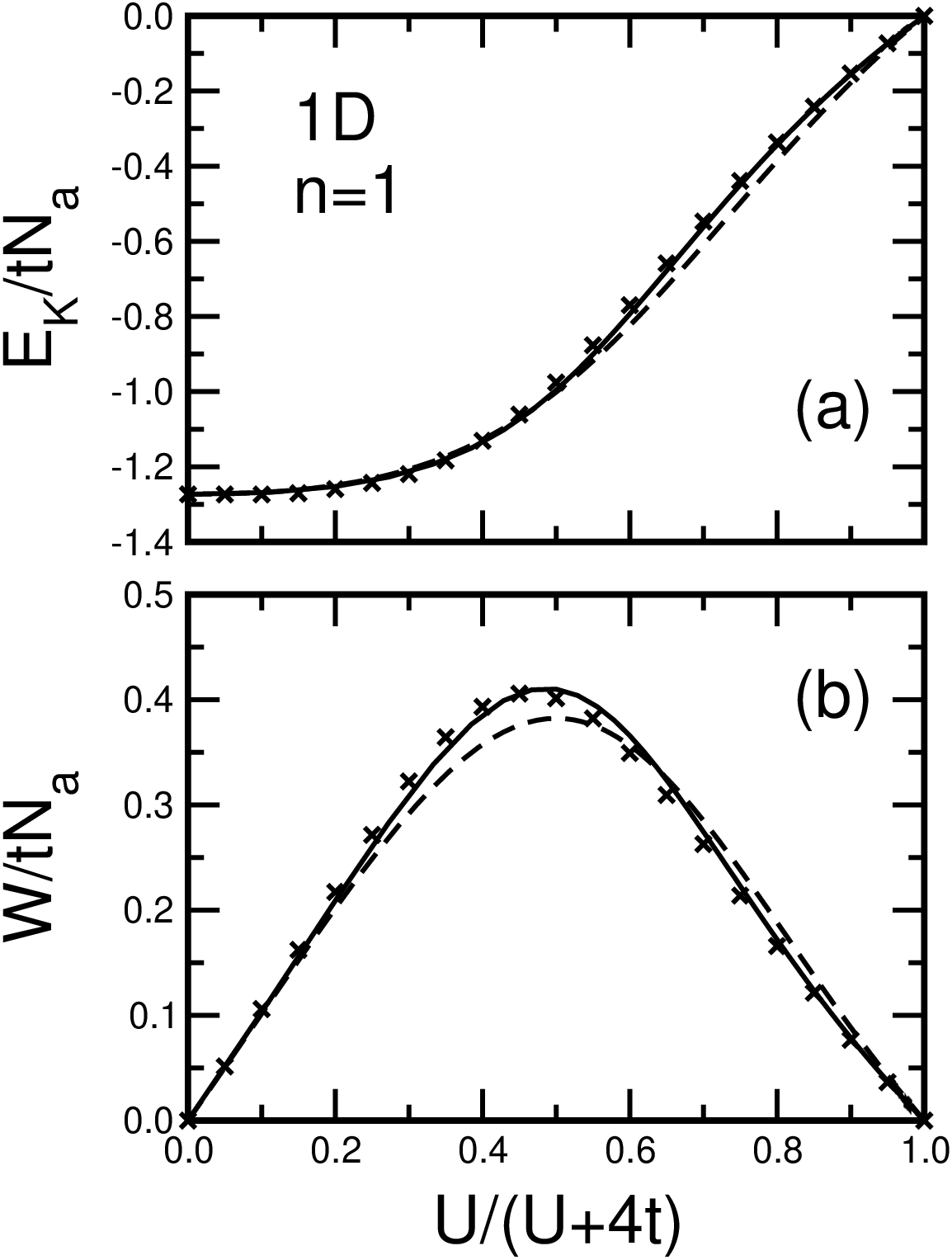}
\caption{\label{fig:EkEc1D}
(a) Kinetic energy $E_{K}$ and (b) Coulomb energy $E_C$ of the 
half-filled 1D Hubbard model as a function of $U/t$. 
The dashed curves correspond to the dimer approximation
[Eq.~(\ref{eq:W2})], the solid curves to the 4th-order approximation 
[Eq.~(\ref{eq:W4}) with $\kappa = \kappa_{1D} = 1.169$],  
and the crosses to the Bethe-ansatz exact solution.\protect\cite{lieb-wu} 
        }
\end{figure}

The higher performance obtained with the 4th-order correction
originates in an improved accuracy of both 
kinetic and Coulomb contributions to the ground-state energy.
As shown in Fig.~\ref{fig:EkEc1D}, 
the kinetic energy $E_K <0$ increases monotonously with increasing 
$U/t$, first rather slowly up to $U/t\simeq 4$, and then more rapidly 
when electron localization starts to set in. 
For $U/t \le 4$, the values of $E_K$ obtained using $W^{(2)}$
and $W^{(4)}$ are very close to the exact result
(typically $|E_K^{(2)} - E_K^{ex}| / E_K^{ex} \le 2.6\%$ and 
$|E_K^{(4)} - E_K^{ex}| / E_K^{ex} \le 2.0\%$). For $U/t > 4$ 
the dimer ansatz shows some limitations while the 4th-order 
approximation remains very accurate 
(for example, for $U/t = 12$, 
$|E_K^{(2)} - E_K^{ex}| / E_K^{ex} \simeq 13\%$ and 
$|E_K^{(4)} - E_K^{ex}| / E_K^{ex} \le 2.4\%$).
The Coulomb energy $E_C$ shows the usual non-monotonous behavior, 
first increasing with $U/t$ in the weakly correlated regime and 
then decreasing as the strongly-correlated limit is approached.
This behavior is correctly described by both 2nd- and 4th-order
approximations. However, one finds that it is in general more 
difficult to accurately describe $E_C$ as compared to $E_K$. 
The 2nd-order $E_C^{(2)}$ underestimates (overestimates)
$E_C^{ex}$ appreciably for $2 < U/t < 5$ ($U/t > 5$).
The 4th-order correction provides a clear improvement over
the dimer ansatz, by increasing $E_C$ in one case ($U/t\le 5$) and 
reducing it in the other ($U/t\ge 10$). 
As for $E_{gs}$, the remaining differences 
with the exact results are quite small and correspond 
to intermediate $U/t$. Summarizing, one may observe 
that the accuracy of the calculated $E_{gs}$ is not the 
result of a strong compensation of errors, since a very 
good performance is achieved for the kinetic and Coulomb energies 
separately. 

\begin{figure}
\includegraphics[scale = 0.4]{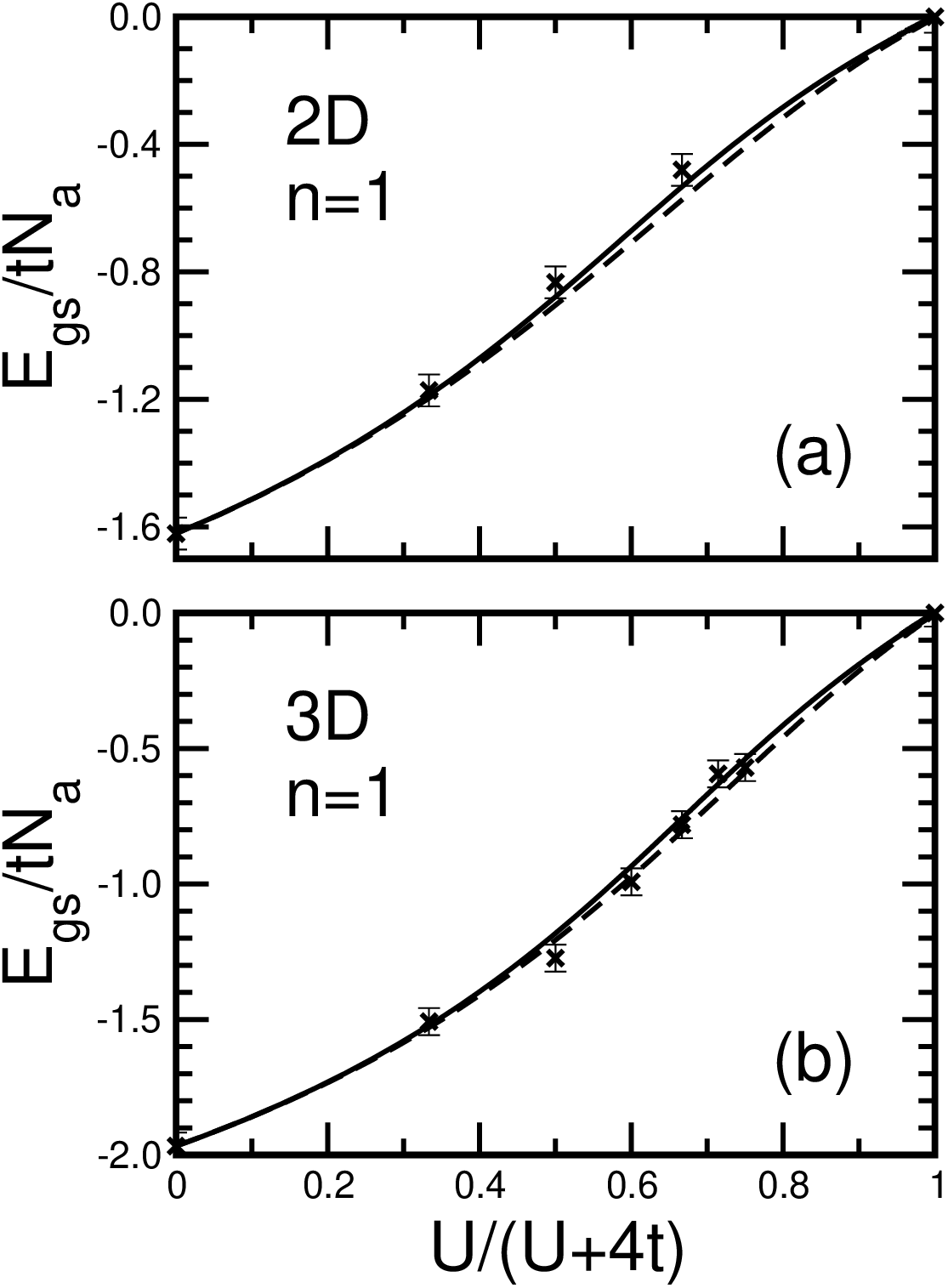}
\caption{\label{fig:Egs2D3D}
Ground-state energy $E_{gs}$ of the half-filled Hubbard model 
as a function of the Coulomb repulsion $U/t$:
(a) two-dimensional (2D) square lattice and 
(b) three-dimensional (3D) simple cubic lattice.
The dashed curves correspond to the dimer approximation
[Eq.~(\ref{eq:W2})] and the solid curves to the 4th-order approximation 
[Eq.~(\ref{eq:W4}) with (a) $\kappa = \kappa_{2D} = 1.135$ and 
(b) $\kappa = \kappa_{3D} = 1.176$]. The crosses with error bars refer to 
quantum Monte Carlo (QMC) calculations.\protect\cite{QMC-2D,QMC-3D}
        }
\end{figure}

In Fig.~\ref{fig:Egs2D3D} results are given for $E_{gs}$ 
of the 2D square lattice and 3D simple cubic lattice at
half-band filling. For $U/t \le 3$ the 2nd-order and 4th-order results
are almost indistinguishable, while for $U/t >4$ the 4th-order 
approximation yields somewhat higher values 
($E_{gs}^{(2)} < E_{gs}^{(4)}  <0$). These trends are very similar
to those observed in the 1D chain. The LDFT calculations for 2D and 3D 
systems compare well with far more demanding quantum Monte 
Carlo (QMC) studies\cite{QMC-2D,QMC-3D} (see Fig.~\ref{fig:Egs2D3D}). 
Furthermore, the reliability of the LDFT results is confirmed by comparison
with exact Lanczos diagonalizations on small clusters, 
for example, on a $N_a= 3 \times 4$ cluster of the square 
lattice with periodic boundary conditions.
In this case, like in 1D, the overall performance is very good, 
with the largest quantitative discrepancies being observed 
for intermediate values of $U/t$. For instance, for $U/t = 4$ one obtains 
$|E_{gs}^{(4)} - E_{gs}^{ex}| / |E_{gs}^{ex}| = 4.2 \times 10^{-2}$, 
and for $U/t=16$ 
$|E_{gs}^{(4)} - E_{gs}^{ex}| / |E_{gs}^{ex}| = 3.2 \times 10^{-2}$.
In conclusion, LDFT using Eq.~(\ref{eq:W4}) for 
the interaction energy $W$ yields an accurate description of the 
ground-state energy of the Hubbard model in different 
dimensions.\cite{foot-comp}

\section{\label{sec:gaps}
Charge excitation gap} 

The charge-excitation or band gap
\begin{equation}
\Delta E_c =  E_{gs}(N_{e}+1) + E_{gs}(N_{e}-1) - 2E_{gs}(N_{e})
\end{equation}
is a property of considerable interest, which measures the
low-energy excitations associated to changes in the 
number electrons $N_e$, and which is very sensitive to the
degree of electronic correlations.
It can be related to the discontinuity 
in the derivative of the ground-state kinetic energy 
$E_K$ and correlation energy $E_{corr} = E_C - E_{\rm HF}$
with respect to band-filling $n$. For $N_a \to \infty$ 
and $n=1$, it is given by 
\begin{equation} 
\Delta E_c = (\partial\varepsilon / \partial n)|_{1^+} -
              (\partial\varepsilon / \partial n)|_{1^-} \; , 
\end{equation} 
where $\varepsilon = (E_K + E_{corr}) / N_a$. 
The determination of $\Delta E_c$ is in general a more difficult
task than the calculation of ground-state properties like $E_{gs}$, 
$E_K$, and $E_C$. In fact, the band-gap in semiconductors has 
been an important problem which motivated numerous works
in the context of DFT in the continuum. Therefore, $\Delta E_c$
appears as a particularly interesting property to investigate
with the present lattice density-functional formalism. 

\begin{figure}
\includegraphics[scale = 0.35]{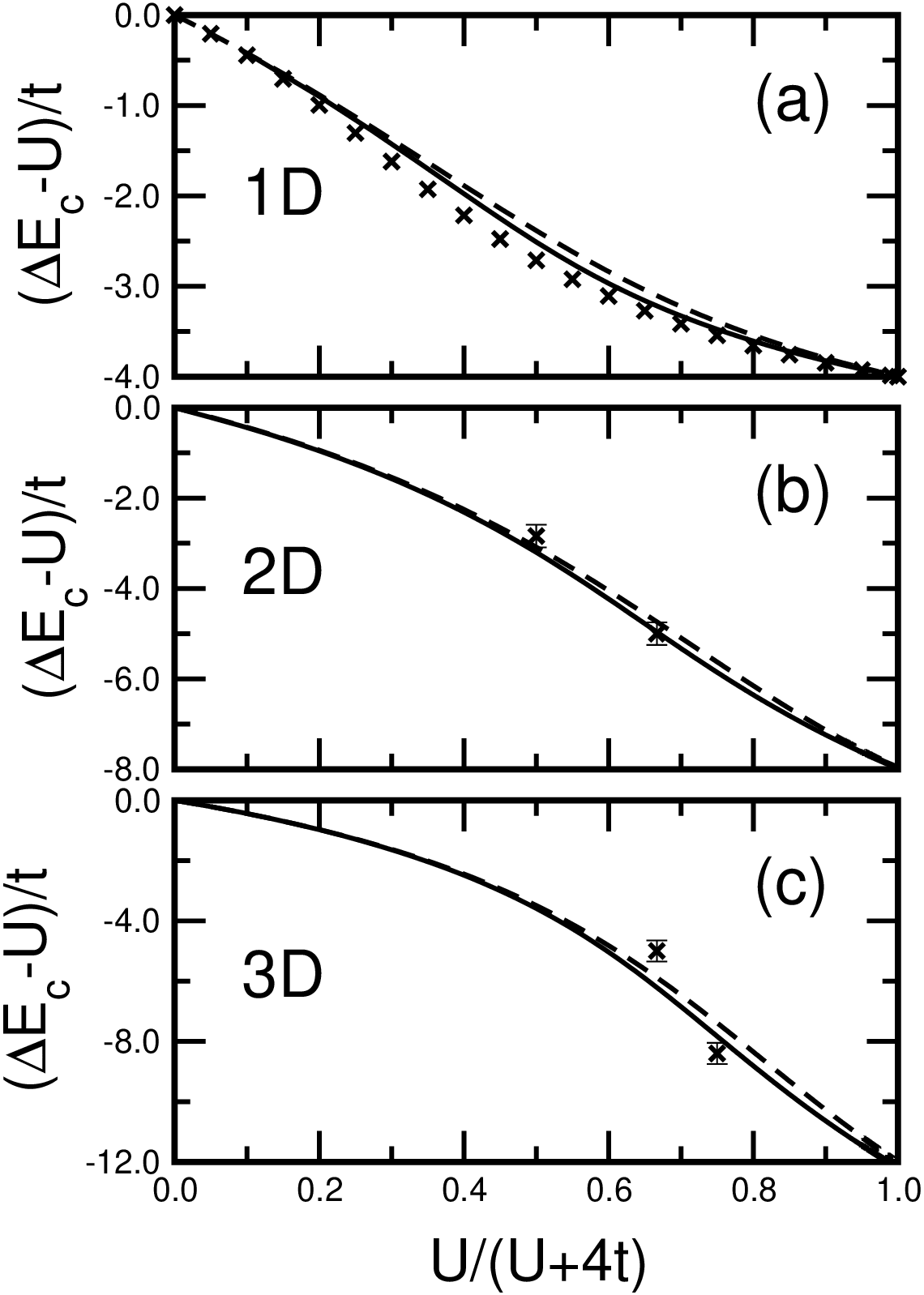}
\caption{\label{fig:gaps}
Charge excitation gap $\Delta E_c$ of the Hubbard model at half-band
filling as a function of $U/t$: 
(a) 1D chain,
(b) 2D square lattice, and 
(c) 3D simple cubic lattice.
The dashed curves correspond to the dimer approximation 
[Eq.~(\ref{eq:W2})] and the solid curves to the 4th-order 
approximation [Eq.~(\ref{eq:W4})] with 
(a) $\kappa = \kappa_{1D} = 1.169$, 
(b) $\kappa = \kappa_{2D} = 1.135$, and 
(c) $\kappa = \kappa_{3D} = 1.176$. 
The crosses refer to exact results in the 1D chain
(Ref.~\protect\onlinecite{lieb-wu}) 
and to QMC calculations in 2D (Ref.~\protect\onlinecite{QMC-2D}) 
and 3D lattices (Ref.~\protect\onlinecite{QMC-3D}). 
        }
\end{figure}

In the half-filled Hubbard model on bipartite lattices, $\Delta E_c$ 
increases with increasing $U/t$ ($\Delta E_c =0$ for $U/t = 0$) and approaches 
the limit $\Delta E_c \to (U - w_b)$ for $U/t \to \infty$, 
where $w_b$ is the width of the single-particle band 
($w_b = 4t$, $8t$, and $12t$ for the 1D, 
2D square, and 3D simple-cubic lattices, respectively). 
Fig.~\ref{fig:gaps} presents LDFT results for $\Delta E_c$ 
in 1D, 2D, and 3D Hubbard models ($n=1$). Comparison with 
the exact Bethe-Ansatz solution for the 1D chain\cite{lieb-wu} and 
with available QMC calculations for the square\cite{QMC-2D} 
and simple cubic\cite{QMC-3D} lattices shows a good overall 
agreement. It should be however noted that in the 1D case 
the gap is significantly overestimated for $U/t\ll 1$. Here we obtain 
$\Delta E_{\rm c} \propto (U/t)^2$, while in the exact solution 
$\Delta E_{\rm c}$ increases much more slowly, namely, 
exponentially in $-t/U$. This discrepancy
concerns both the 2nd-order and the 4th-order approximations,
which are nearly indistinguishable for $U/t < 2$--$4$.
Consequently, it is possible that the results for 2D and 3D lattices
shown in Fig.~\ref{fig:gaps} also overestimate the gap for small $U/t$.
In any case, the accuracy of LDFT improves rapidly with increasing $U/t$,
as electron localization starts to set in, and the error 
in $\Delta E_{\rm c}$ tends to zero for large $U/t$. Therefore, the 
development of a Mott insulator in the strongly correlated limit is 
correctly described. 

It is important to remark, in the context of metal-insulator transitions 
in three dimensions, that our calculations on the SC lattice yield
a finite gap $\Delta E_c >0$ for $n=1$ and all $U/t >0$. 
This is consistent with 
previous results on 3D bipartite lattices, which are expected to be 
antiferromagnetic (AF) insulators for all $U/t>0$.\cite{georges,imada} 
The functionals 
$W^{(2)}$ and $W^{(4)}$ reproduce correctly this behavior, as well 
as the formation of local moments $\langle S_i^2 \rangle = 
3(1 - 2 \langle \hat n_{i\uparrow} \hat n_{i\downarrow}\rangle)/4$, without 
involving a spin-density-wave symmetry breaking. This can be understood be 
recalling that they are based on the exact functional of the Hubbard dimer 
which, being a bipartite cluster,
incorporates AF correlations ($n=1$). However, the properties change 
qualitatively if frustrations become important (e.g., in non-bipartite 
lattices or if second NN hopping are significantly large). In this case 
it has been shown that the half-filled Hubbard model
is a metal with $\Delta E_c =0$
for small $U>0$ and that a metal-insulator transition takes place at a finite 
interaction strength $U_c$, which is of the order of the single-particle band 
width $w_b$.\cite{georges} This behavior is not reproduced by 
the functionals $W^{(2)}$ and $W^{(4)}$, even if they are applied to
compact lattices (e.g., the face-centered cubic lattice), since they are 
free from any singularities throughout the domain of representability 
(except for $g_{12} = 1$) and since the resulting $\gamma^{gs}_{ij}$
are smooth functions of $U/t$. Notice that the exact 
functional $W_{ex}$ may show a far more complex behavior, particularly
if the nature of the state $|\Psi[\gamma] \rangle$ yielding the 
minimum of Eq.~(\ref{eq:Wex}) changes as a function of $\gamma$. 
This is expected to be the case at a metal-insulator transition, where 
a discontinuous decrease of 
$\langle \hat n_{i\uparrow} \hat n_{i\downarrow}\rangle$ occurs.
Finally, let us point out that the results derived from
Eqs.~(\ref{eq:W2}) and (\ref{eq:W4}) for large $U/t$ ($U > U_c$)
are consistent with previous studies. This concerns not only the 
presence of a finite gap $\Delta E_c$, but also the fact that the number 
of double occupations does not vanish on the insulating side of the 
transition.\cite{georges,imada}

Comparing 2nd- and 4th-order approximations 
for $U/t > 2$--$4$ one observes that the charge gap is always 
somewhat smaller in the latter case. For the 1D chain, the 
reduction of $\Delta E_{\rm c}$ due to the 4th-order correction
improves the agreement with the exact solution appreciably 
(e.g., $|\Delta E_{\rm c}^{(4)} - \Delta E_{\rm c}^{ex}| \simeq
  |\Delta E_{\rm c}^{(2)} - \Delta E_{\rm c}^{ex}| /2$ for $U/t>10$). 
In the considered 2D and 3D lattices, the differences 
between 2nd- and 4th-order results are similar to those observed
in the 1D chain. Comparison with QMC calculations shows a good overall 
agreement although some quantitative differences can be noted. 
For example, as shown in Fig.~\ref{fig:gaps}, 
our values for $\Delta E_{\rm c}$ are somewhat smaller than 
the QMC ones for the 2D (3D) lattice with $U/t = 4$ ($U/t = 8$)
and somewhat larger for $U/t = 8$ ($U/t = 12$). 
In summary, the ensemble of 1D, 2D and 3D results shows 
that LDFT provides a very simple and efficient method of 
calculating the charge excitation energies of the Hubbard model 
in different dimensions and interaction regimes. However, 
the proposed approximations to $W$ are still not quite satisfactory
in the weakly-correlated limit and deserve to be improved.

\section{\label{sec:chsus}
Charge susceptibility} 

\begin{figure}
\includegraphics[scale=0.4]{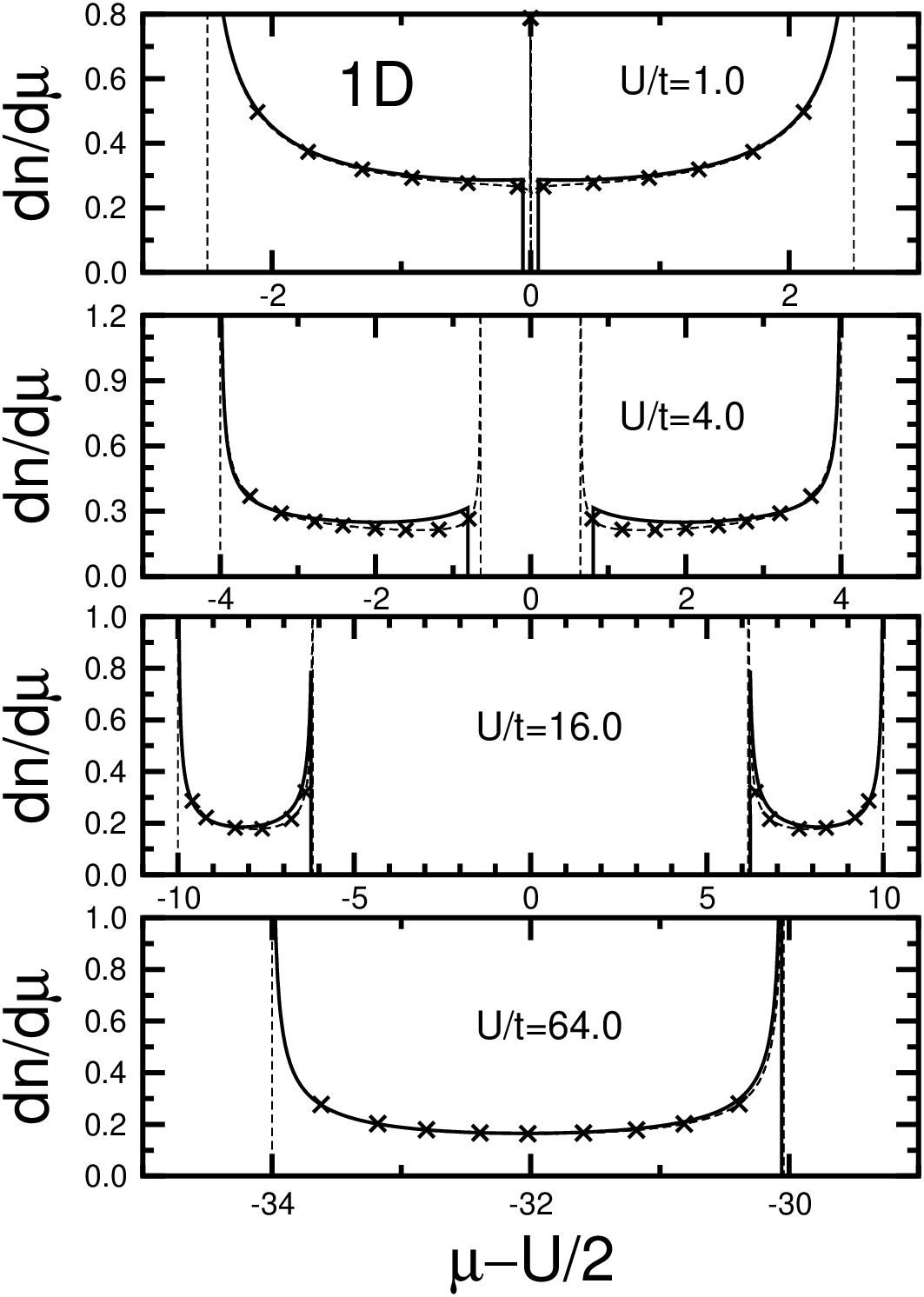} 
\caption{\label{fig:dos1D}
Charge susceptibility $\chi_{\rm c}$ of the 1D Hubbard model 
as a function of the chemical potential $\mu$ for different Coulomb 
repulsions $U/t$. The solid curves refer to LDFT in the dimer 
approximation [Eq.~(\ref{eq:W2})] and the dashed curves with crosses 
to the exact Bethe ansatz solution.\protect\cite{lieb-wu} 
For $U/t=64$ only the lower Hubbard band is shown.
        }
\end{figure}

\begin{figure}
\includegraphics[scale=0.4]{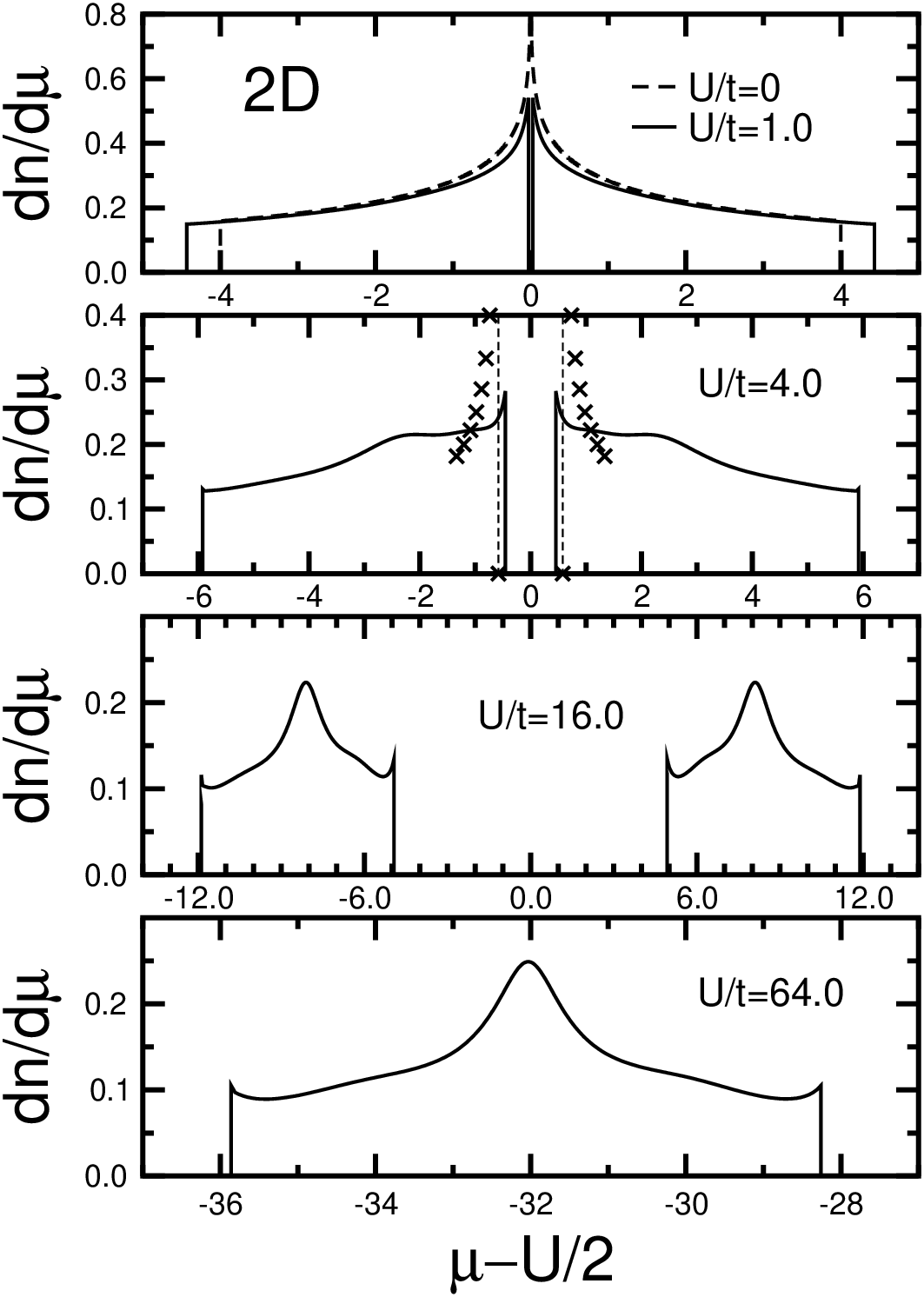} 
\caption{\label{fig:dos2D}
Charge susceptibility $\chi_{\rm c}$ of the Hubbard model on
a 2D square lattice as a function of the chemical potential 
$\mu$ for different Coulomb repulsions $U/t$. 
The solid curves are obtained using LDFT and the dimer 
approximation to the interaction-energy functional [Eq.~(\ref{eq:W2})].
The crosses for $U/t = 4$ refer to ground-state QMC calculations 
(Ref.~\protect\onlinecite{QMC-2D}).
For $U/t=64$ only the lower Hubbard band is shown.
        }
\end{figure}

\begin{figure}
\includegraphics[scale=0.4]{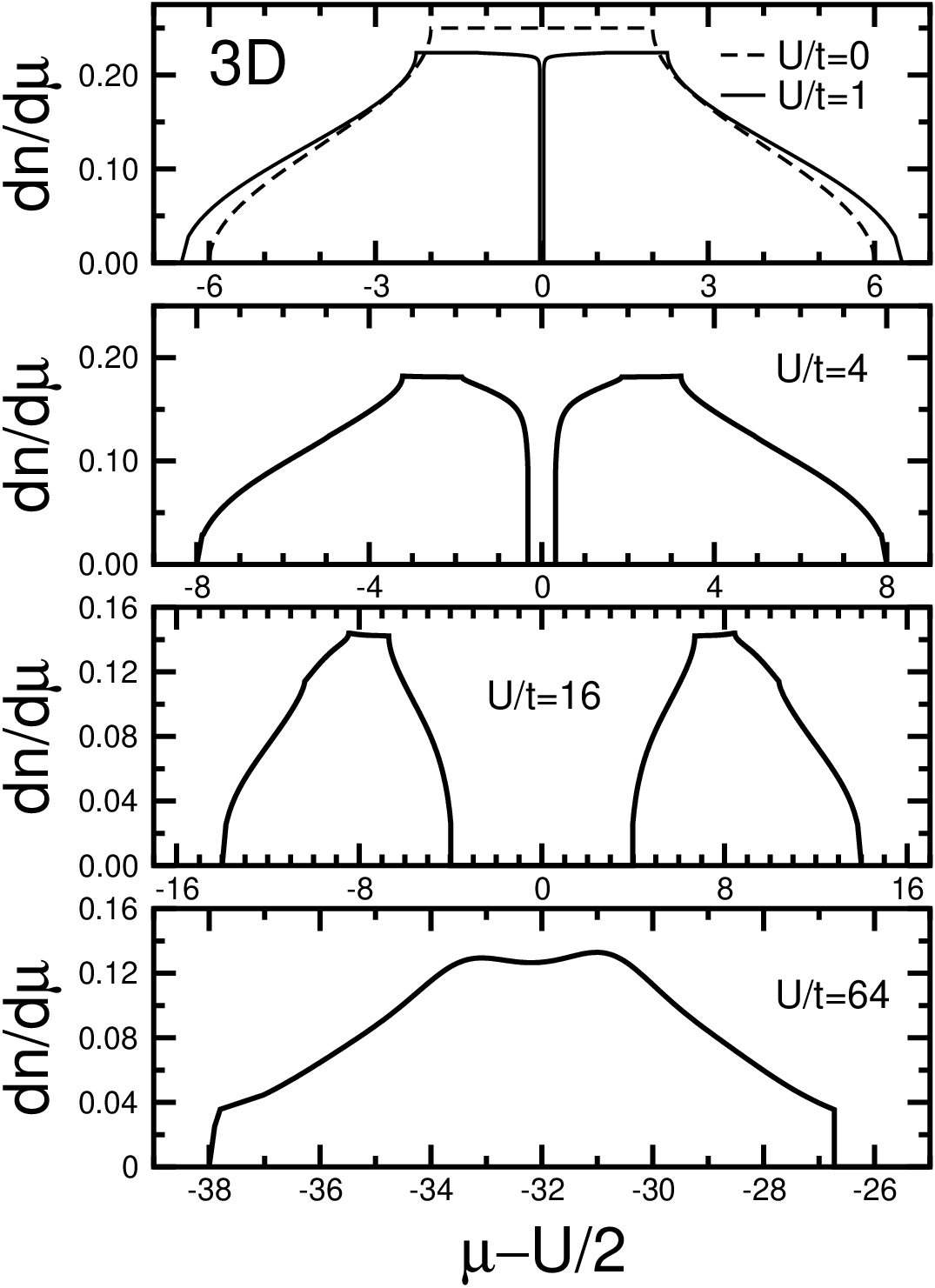} 
\caption{\label{fig:dos3D}
Charge susceptibility $\chi_{\rm c}$ of the Hubbard model on
a 3D simple cubic lattice as a function of the chemical potential 
$\mu$ for different Coulomb repulsions $U/t$. 
The results are obtained using the dimer 
approximation to the interaction-energy functional [Eq.~(\ref{eq:W2})].
For $U/t=64$ only the lower Hubbard band is shown.
        }
\end{figure}

The charge susceptibility $\chi_c$ is defined by 
\begin{equation}
\chi_c = \frac{d n} {d \mu} \; ,
\end{equation}
where $n=N_e/N_a$ is the number of electrons per site 
and $\mu$ the chemical potential. It represents the 
many-body density of electronic states at the Fermi energy $\mu$  
and thus provides very useful information on the low-energy 
charge-excitation spectrum as a function of band filling.
In Figs.~\ref{fig:dos1D}--\ref{fig:dos3D} $\chi_c$ is given 
as a function of $\mu$ for 1D, 2D, and 3D Hubbard models
on bipartite lattices for representative values of $U/t$. 
The LDFT calculations reported in these figures were performed using
the dimer ansatz for $W$ given by Eq.~(\ref{eq:W2}). 
As it will be discussed below, the 4th-order approximation 
[Eq.~(\ref{eq:W4})] yields very similar results. 
In the case of the 1D chain comparison is made with the exact $\chi_c(\mu)$,
which is obtained from the Bethe ansatz solution.\cite{lieb-wu} 
For the 2D square lattice, we also show ground-state QMC 
results\cite{QMC-2D} for $U/t = 4$.
Notice that in bipartite lattices, as those considered here, electron-hole 
symmetry implies that $\chi_c$ is the same for band fillings $n$
and $n'=2-n$, and therefore $\chi_c(\mu) = \chi_c(\mu'= U - \mu)$. 


In the absence of interactions $\chi_c$ coincides 
with the single-particle density of states of the corresponding 
lattices ($U=0$). These are gapless 
and show the usual van Hove singularities at the band edges 
$\mu = \pm w_b/2$ and at some points within the bands of the
square and simple-cubic lattices.
For finite $U$ a gap $\Delta E_c = \mu({n=1^+}) - \mu({n=1^-})$ 
opens at half-band filling which increases monotonously with $U$,
as discussed in the previous section. Thus, the two so-called 
lower and upper Hubbard bands start to be distinguished,
which correspond to hole and electron doping respectively. 
The separation of the bands becomes particularly clear for $U\approx w_b$,
when $\Delta E_c$ reaches values of the order of single-particle band 
width $w_b$ (see Figs.~\ref{fig:dos1D}--\ref{fig:dos3D}).
At the same time the width of the lower and upper bands increases
with $U$, from $w_b/2$ 
for $U=0^+$, to $w_b$ 
for $U=+\infty$. These qualitative features are common to 
bipartite lattices in all dimensions
and are correctly described by LDFT. 

In the 1D case, where 
a detailed comparison with the exact solution is possible, 
we observe that our results are very accurate except close to 
half-band filling and for small or moderate values of $U/t$
(see Fig.~\ref{fig:dos1D}). 
The nature of the discrepancies close to $n=1$ is basically twofold.
First, we find again the overestimation of the 
band gap $\Delta E_c$ which is relatively important 
for small $U/t$ (see also Sec.~\ref{sec:gaps}). 
Consequently, the band edges $\mu({n=1^-})$ and
$\mu({n=1^+})$ are not precisely reproduced in this limit, 
even if the absolute error 
$\epsilon = |\mu - \mu^{ex}|$
always remains reasonably small. 
The largest inaccuracies are found for $U/t \simeq 3$ and amount to 
$\epsilon / w_b = 8.1\times 10^{-2}$. Nevertheless, this problem 
disappears as $U/t$ increases, since $\epsilon$ tends rapidly 
to zero in the strongly correlated limit 
(e.g., $\epsilon / w_b = 1.2 \times 10^{-4}$ for $U/t = 16$). 
The second limitation concerns the shape of 
$\chi_c$ close to half-band filling.   
The exact solution of the 1D chain shows sharp 
divergences in $\chi_c$ at the gap edges $\mu({n=1^-})$ 
and $\mu({n=1^+})$ for $U>0$, which we fail to reproduce.  
For small and moderate $U/t$, for example $U/t = 1$ or $4$, 
we obtain a nearly constant $\chi_c$ for $\mu \to \mu(n=1^\pm)$, 
while the exact result is $\chi_c \to +\infty$ (see Fig.~\ref{fig:dos1D}).
Notice, however, that the increase and divergence of $\chi_c^{ex}$ 
are sharply localized in a narrow range of $\mu$, particularly for
small $U/t$. The divergence of $\chi_c$ for $n\to 1$ could be 
reproduced by considering broken symmetry solutions of the LDFT equations, 
like in the AF Hartree-Fock approximation. Even so, it 
would be more interesting to describe this effect
without involving a symmetry breaking, which is known to be artificial,
and which could affect the results on the kinetic, 
Coulomb, and total energies, particularly in the case of finite 
systems.\cite{maol} As we approach the strongly 
correlated limit the LDFT results for $\chi_c$ develop peaks at the 
gap edges, which height increases with $U/t$, thus approaching 
asymptotically the exact result. Still, $\chi_c$ always remains 
finite for all finite $U/t$ (see Fig.~\ref{fig:dos1D} for 
$U/t = 16$ and $64$).
The very good performance for large $U/t$ can be understood by 
recalling that for $U/t = +\infty$, the LDFT results correspond to 
the fully-polarized or Nagaoka state,\cite{nagaoka} 
which is the exact ground-state in 1D 
for all $n$ ($U/t = +\infty$).\cite{lieb-wu}
In higher dimensions it is possible that our calculations
yield a finite-height peak for $n\to 1$,
where a true divergence of $\chi_c$ could be present. 
This seems to be the case in the 2D square lattice where we observe 
narrow peaks in $\chi_c$ at the band edges.
In fact ground-state QMC calculations on the square lattice with $U/t=4$ 
predict a divergent $\chi_c$ at half-band filling (see Fig.~\ref{fig:dos2D}).
In contrast, no such peaks are found in our calculations of $\chi_c$
for the 3D simple-cubic lattice (see Fig.~\ref{fig:dos3D}).

As already discussed in the previous section, it is important to 
remark that the results presented in Fig.~\ref{fig:dos3D} are 
representative of bipartite lattices which at the half-band filling
show an AF insulating behavior for all $U/t >0$. In this case of our 
results are in good qualitative agreement with previous studies. 
The obtained simple Hubbard-approximation-like structure of $\chi_c$, 
with a lower and upper Hubbard bands, also applies to frustrated lattices 
or to paramagnetic phases provided that $U/t$ is sufficiently large to 
bring the system on the insulating side of the metal-insulator 
transition ($U>U_c$).\cite{georges,imada} However, it has been shown 
that the presence of frustrations 
drives the system into an AF {\em metallic} state at small 
$U/t$ which contrasts with the AF insulator found in the absence of 
frustrations.\cite{georges} In this case the spectral density presents
---in addition to a progressive development of lower and upper Hubbard bands
with increasing $U/t$--- a Kondo-like resonance at half-band filling 
($\mu-U/2=0$) characterized by a constant-height peak having a width that 
decreases with increasing $U/t$ and that vanishes at $U_c$, i.e., at the 
transition to the insulating state (see Ref.~\onlinecite{georges}). 
The functionals $W^{(2)}$ or $W^{(4)}$ fail to reproduce this kind 
of behavior, even when applied to compact structures (e.g., fcc lattice). 
This limitation does not seem surprising, 
since Eqs.~(\ref{eq:W2}) and (\ref{eq:W4}) 
were derived from the properties of a bipartite system, and since the
extensions presented in this paper, while achieving an accurate $W$ 
in the strongly correlated limit at half-band filling, do not aim
a precise description at small $U/t$ and as a function of $n$.
Exploring the functional dependence of $W$ in frustrated structures,
particularly close to $n=1$ and $g_{12}\simeq 1$, should 
provide very useful clues in view of developing practical approximations
capable of describing these remarkable effects.

Let us finally point out that we have also determined $\chi_c$ using 
$W^{(4)}$ as approximate interaction energy [Eq.~(\ref{eq:W4})]
and found that the results are 
very similar to those obtained using $W^{(2)}$ and shown 
in Figs.~\ref{fig:dos1D}--\ref{fig:dos3D}. In both cases the 
results are extremely good away from half-band filling,
nearly indistinguishable from the 1D exact solution. 
Close to $n=1$, the 4th-order calculations yield smaller $\Delta E_c$ and 
thus perform slightly better for $U/t\le 4$. However, the 
divergences of $\chi_c$ at the gap edges are not reproduced. 
Therefore, the 4th-order corrections do not provide a significant 
improvement over the dimer ansatz concerning $\chi_c$ of the 1D chain. 
This is probably related to the simple form considered for $W^{(4)}$ 
which uses a coefficient $\kappa$ that is independent of $n$ 
[see Eq.~(\ref{eq:W4})]. While this approximation seems satisfactory for 
applications that concern a fixed band filling, it appears 
as a limitation for properties like $\Delta E_{\rm c}$ or 
$\chi_c$, where a precise description of the dependence of 
$W$ on $n$ is crucial. For 2D and 3D lattices the 
4th-order results for $\chi_c$ are also very similar
to those shown in Figs.~\ref{fig:dos2D} and \ref{fig:dos3D}.

\section{\label{sec:concl}
Discussion} 

A new approximation to the interaction-energy functional 
$W[\gamma]$ of the Hubbard model has been proposed in the
framework of lattice density functional theory, which 
exactly recovers the limit of strong electron correlations 
at half-band filling. The simpler ansatz which was derived from 
the functional dependence of $W$ in the Hubbard dimer\cite{ldftscfer}
is thereby extended and improved. A more accurate 
description of $W$ is achieved throughout the domain 
of representability of $\gamma_{ij}$ including the
crossover from weak to strong correlations.
Several properties have been determined by applying this functional 
to one-, two-, and three-dimensional lattices. Ground state energies,
as well as kinetic and Coulomb energies, were successfully determined 
in all dimensions and interaction regimes. 
Very good results are also obtained concerning the charge-excitation gap 
$\Delta E_c$ and the charge susceptibility $\chi_c$ of bipartite lattices,
except very close to half-band filling ($n=1$) and for small values 
of the Coulomb repulsion strength ($U/t\le 4$). This reveals some
limitations in the description of the band-filling dependence of $W$
for $n\simeq 1$ and $\gamma_{12} \simeq \gamma_{12}^0$, which 
deserve more detailed investigations. Further insight on the origin of
this problem could be obtained, for example, by analyzing the properties of
the exact $W$ as derived from the Bethe ansatz exact solution of the 1D
chain and from Lanczos diagonalizations in finite 2D clusters with 
periodic boundary conditions. Moreover, the functional dependence 
of $W$ could be determined in the limit of large $\gamma_{12}$ (i.e., 
$\gamma_{12} \to \gamma_{12}^0$ corresponding to the weak correlations)
by applying perturbation theory for small $U/t$. In this way, more 
accurate approximations to $W$ could be developed in order to improve the
results on $\Delta E_c$ and $\chi_c$ in this limit, particularly concerning
the differences between bipartite and non-bipartite lattices.

Besides these methodological aspects, the accuracy of the
results and the simplicity of the calculations encourage 
new applications of the present approach to related problems
of current interest like dimerized one-dimensional chains and 
ladders, the 2D square lattice with competing first and second 
nearest-neighbor hoppings, the properties of $\pi$ electrons
in doped fullerenes and nanotubes in the framework of Hubbard or PPP 
models, or the connection with the continuum's DFT using minimal basis 
sets. In this way, a novel density-functional route to the physics 
of strongly correlated fermions is opened.

\begin{acknowledgments}

This work has been supported by CONACyT Mexico through 
Grant No.\ J-41452 and Millennium Initiative W-8001.
Computer resources were provided by IDRIS (CNRS, France).

\end{acknowledgments}


\begin{thebibliography}{}
\bibitem {foot_ens}
Ensemble density-matrices are written as $\Gamma_{ij} = \sum_n w_n 
\langle\Psi_n| \sum_\sigma \hat c_{i \sigma }^{\dagger} \hat c_{j \sigma }
|\Psi_n \rangle$ with $w_n \ge 0$ and $\sum_n w_n = 1$. 
In practice, they are much easier to characterize than 
pure-state density matrices.
\bibitem{foot_gen}
The LDFT formalism can be easily extended to arbitrary interactions
$H_I = (1/2) \sum V_{ijkl} 
\hat c_{i \sigma }^{\dagger} \hat c_{k   \sigma '}^{\dagger} 
\hat c_{l \sigma '} \hat c_{j \sigma}$, by replacing 
$U \sum_i \hat n_{i\uparrow} \hat n_{i\downarrow}$ by $H_I$ in 
the expressions for the interaction-energy functional $W[\gamma]$.
However, note that the functional $W[\gamma]$ depends crucially 
on the form of $V_{ijkl}$. See also Refs.~\protect\onlinecite{ldftxcfun} 
and \protect\onlinecite{ldftdim1D}.
\bibitem{nonbip}
For simplicity we focus here on bipartite lattices choosing 
$\gamma_{12}>0$ and $t_{ij} = -t<0$ for NN $ij$. 
Non-bipartite lattices can be treated 
analogously by considering positive 
and negative domains of $\gamma_{12}$ separately, 
positive (negative) $\gamma_{12}$ being relevant for 
negative (positive) values of the hopping integrals $t_{ij}$
[see Eq.~(\ref{eq:hamhub})]. As shown in Ref.~\onlinecite{ldftxcfun},
the scaling behavior of $W / E_{\rm HF}$ as a function of 
$g_{12} = (\gamma_{12}      - \gamma_{12}^{\infty-}) / 
          (\gamma_{12}^{0-} - \gamma_{12}^{\infty-})$
is also observed for 
$\gamma_{12}^{0-} \le \gamma_{12} \le \gamma_{12}^{\infty -} \le 0$.
A similar situation occurs in dimerized systems where the 
domain of representability 
$[\gamma^\infty(\phi), \gamma^0(\phi)]$ depends 
on the ratio $\gamma_{12}/\gamma_{23}$ between 
the bond orders corresponding to short and long bonds
[$\phi=\arctan(\gamma_{12}/\gamma_{23})$, see Ref.~\onlinecite{ldftdim1D}].
\bibitem{foot-comp}
Concerning the computational efficiency of LDFT, one may notice
that the determination of $E_{gs}$, $\gamma_{ij}^{gs}$, and other derived 
properties involves very simply analytical expressions which correspond 
to the minimization of $E = E_K + W$ with respect to $\gamma_{ij}$
[see Eq.~(\protect\ref{eq:tij})]. In the case of periodic systems these 
can be solved straightforwardly for any value of $U/t$ and $n$. Most of 
the actually modest calculation effort goes in the determination of the 
functionals $W^{(2)}$ or $W^{(4)}$. This requires calculating 
$\gamma_{12}^0$ and $\gamma_{12}^\infty$ only once for each system
in order to define the domain of representability of $\gamma_{12}$. 
These preliminary calculations, as well as the number of variables
involved in the minimization of $E$, become of course increasingly 
important as the symmetry of the systems is lowered (see, for example, 
Ref.~\protect\onlinecite{ldftdim1D}).
\end{thebibliography}
\end{document}